\pdfoutput=1

\documentclass[11pt,twoside,a4paper,cmspaper,final,collab]{cms-tdr}
\def\svnVersion{404129}\def\cmsCernNoTag{CERN-EP/2016-249}\def\cmsCernDate{\today}\def\cmsMessage{Published in Physics Letters B as \href{http://dx.doi.org/10.1016/j.physletb.2017.04.028}{\doi{10.1016/j.physletb.2017.04.028.}}}

\begin{document}\cmsNoteHeader{TOP-12-031}

\hyphenation{had-ron-i-za-tion}
\hyphenation{cal-or-i-me-ter}
\hyphenation{de-vices}
\RCS$Revision: 404018 $
\RCS$HeadURL: svn+ssh://svn.cern.ch/reps/tdr2/papers/TOP-12-031/trunk/TOP-12-031.tex $
\RCS$Id: TOP-12-031.tex 404018 2017-05-11 09:47:02Z stiegerb $
\newlength\cmsFigWidth
\ifthenelse{\boolean{cms@external}}{\setlength\cmsFigWidth{0.85\columnwidth}}{\setlength\cmsFigWidth{0.4\textwidth}}
\ifthenelse{\boolean{cms@external}}{\providecommand{\cmsLeft}{top}}{\providecommand{\cmsLeft}{left}}
\ifthenelse{\boolean{cms@external}}{\providecommand{\cmsRight}{bottom}}{\providecommand{\cmsRight}{right}}
\cmsNoteHeader{TOP-12-031}

\newcommand{\x}{\ensuremath{\phantom{0}}}

\title{Measurement of the mass difference between top quark and antiquark in pp collisions at \texorpdfstring{$\sqrt{s} = 8\TeV$}{sqrt(s) = 8 TeV}}

\date{\today}
\abstract{
The invariance of the standard model (SM) under the CPT transformation predicts equality of particle and antiparticle masses. This prediction is tested by measuring the mass difference between the top quark and antiquark ($\Delta m_{\PQt} = m_{\PQt} - m_{\PAQt}$) that are produced in pp collisions at a center-of-mass energy of 8\TeV, using events with a muon or an electron and at least four jets in the final state. The analysis is based on data corresponding to an integrated luminosity of 19.6\fbinv collected by the CMS experiment at the LHC, and yields a value of $\Delta m_{\PQt} = -0.15 \pm 0.19\stat \pm 0.09\syst\GeV$, which is consistent with the SM expectation. This result is significantly more precise than previously reported measurements.
}
\hypersetup{
pdfauthor={CMS Collaboration},
pdftitle={Measurement of the mass difference between top quark and top antiquark in pp collisions at 8 TeV},
pdfsubject={CMS},
pdfkeywords={CMS, top quark physics}
}
\maketitle
\section{Introduction}
\label{sec:intro}
Symmetries such as charge conjugation (C), parity or space reflection (P), and time reversal (T) play a fundamental role in the standard model (SM) of particle physics~\cite{SM_glashow,SM_weinberg,SM_salam}.
The invariance of the SM under their combination, the CPT symmetry, predicts equality of particle and antiparticle masses, and thus far experiments have confirmed this prediction~\cite{PDG}.
In some extensions of the SM, however, CPT-violating effects are present~\cite{PhysRevD.55.6760,Barenboim200373,Raghavan:2003du,ALANKOSTELECKY1991545}.
The large number of top quarks produced in proton-proton ($\Pp\Pp$) collisions at the CERN LHC provides an opportunity to test CPT symmetry in the quark sector for the most massive particle of the SM through a precise measurement of the difference in mass between the top quark (\cPqt) and its antiparticle (\cPaqt), $\Delta m_{\cPqt} \equiv m_{\cPqt} - m_{\cPaqt}$.
This quantity has been measured previously in $\Pp\Pap$ collisions at $\sqrt{s}$ = 1.96\TeV\ by the CDF and D0 experiments, and in pp collisions at $\sqrt{s} = 7\TeV$ by the ATLAS and CMS experiments.
The CDF measurement of $\Delta m_{\cPqt} = -3.3 \pm 1.4\stat \pm 1.0\syst\GeV$~\cite{CDFmassDiff} is almost two standard deviations away from the SM value.
However, the subsequent measurements by D0, CMS, CDF, and ATLAS,
yielding $0.8 \pm 1.8\stat \pm 0.5\syst\GeV$~\cite{D0massDiff},
$-0.44 \pm 0.46\stat \pm 0.27\syst\GeV$~\cite{CMStopMassDiff},
$-1.95 \pm 1.11\stat \pm 0.59\syst\GeV$~\cite{Aaltonen:2012zb},
and $0.67 \pm 0.61\stat \pm 0.41\syst\GeV$~\cite{ATLASmassDiff}, 
respectively, are all in agreement with CPT symmetry.
This article presents a measurement performed with the CMS~\cite{CMS:2008zzk} detector at the LHC.\@
It represents the first determination of $\Delta m_{\cPqt}$ at $\sqrt{s} = 8\TeV$.
While the same techniques as for the previous $\Delta m_{\cPqt}$ measurement by CMS~\cite{CMStopMassDiff} are used, both the statistical and the systematic uncertainty are significantly reduced.
The event selection is optimized for \ttbar\ production where one of the \PW\ bosons decays hadronically ($\cPqt \to \cPqb\PWp \to \cPqb\cPq\cPaq'$, or its charge conjugate) and the other decays leptonically ($\cPqt \to \cPqb\PWp \to \cPqb\ell^+\nu_{\ell}$, or its charge conjugate), with $\ell$ corresponding to either an electron or a muon (including also decays to $\tau$ leptons where the $\tau$ decays leptonically).
The data are split into $\ell^-$ and $\ell^+$ samples that contain three-jet decays of the associated top quarks or antiquarks, respectively. For each event category, the ideogram likelihood method \cite{Abdallah:2008xh} is used to measure $m_{\cPqt}$ and $m_{\cPaqt}$, from which their difference is obtained.
\section{The CMS detector}
\label{sec:CMS}
The central feature of the CMS apparatus is a superconducting solenoid of 6\unit{m} internal diameter, providing a magnetic field of 3.8\unit{T}. The field volume houses a silicon pixel and strip tracker, a crystal electromagnetic calorimeter (ECAL), and a brass/scintillator hadron calorimeter (HCAL). The inner tracker reconstructs charged-particle trajectories within the pseudorapidity range $|\eta| < 2.5$.
The tracker provides an impact parameter resolution of 20--30\unit{$\mu$m} and a resolution of the momentum transverse to the beam direction ($\pt$) of 1--3\% for 10\GeV\ charged particles.
Muons are measured for $|\eta|< 2.4$ using detection planes based on three technologies: drift tubes, cathode-strip chambers, and resistive plate chambers. Matching outer muon trajectories to tracks measured in the silicon tracker provides a \pt\ resolution of 1--6\% for the $\pt$ values relevant to this analysis~\cite{MuonPerformance}.
In the region of $\vert \eta \vert< 1.74$, the HCAL cells have widths of 0.087 in pseudorapidity and 0.087\unit{rad} in azimuth ($\phi$).
In the $(\eta,\phi)$ plane, for $\vert \eta \vert< 1.48$, the HCAL cells map onto arrays of $5 \times 5$ ECAL crystals to form calorimeter towers that project radially outwards from near the center of the CMS detector. At larger values of $\vert \eta \vert$, the size of the towers increases and the matching ECAL arrays contain fewer crystals.
The energy resolution is less than $5\%$ for the electron energies considered in this analysis~\cite{ECALPerformance,Khachatryan:2015hwa}.
In addition to the barrel and endcap detectors, CMS has extensive forward calorimetry.
A more detailed description of the CMS detector, together with a definition of the coordinate system used and the relevant kinematic variables, can be found in Ref.~\cite{CMS:2008zzk}.
\section{Data and simulation}
\label{sec:samples}
The data used in this analysis correspond to an integrated luminosity of $19.6\pm0.5\fbinv$~\cite{CMS-PAS-LUM-13-001}, collected during the 2012 $\Pp\Pp$~collision run of the LHC at a center-of-mass energy of 8\TeV.
Events are selected online using a trigger that requires an isolated electron with $\pt>27\GeV$ and $|\eta|<2.5$ or an isolated muon with $\pt>24\GeV$ and $|\eta|<2.1$.
Simulated Monte Carlo (MC) samples of \ttbar, \PW, and \Z\ boson production are generated with \MADGRAPH\ 5.1.3.30~\cite{MadGraph} interfaced with \PYTHIA\ 6.4.26~\cite{pythia} for parton showering.
Single top quark events are simulated using the \POWHEG\ generator~\cite{Frixione:2007vw,Alioli:2010xd,Alioli:2009je,Re:2010bp}, also interfaced to \PYTHIA.\@
For studies of systematic effects, a sample of \ttbar\ events is generated with \textsc{mc@nlo} 3.14~\cite{MCatNLO}, combined with \HERWIG\ 6.520~\cite{Herwig} for parton showering.
All generated events are passed through a simulation of the CMS detector based on \GEANTfour~\cite{Geant4}. The simulation includes the effect of pileup, \ie\ additional $\Pp\Pp$~collisions occurring during the same bunch-crossing or immediately preceding or following the primary crossing.
The theoretical cross sections for \PW\ boson and \Z\ boson production are calculated with \textsc{FEWZ}~\cite{FEWZ} at next-to-next-to-leading-order (NNLO) precision of quantum chromodynamics (QCD).
The \ttbar\ and single top quark production cross sections are from calculations at NNLO~\cite{ttXS-NNLO} and approximate NNLO~\cite{Kidonakis} precision, respectively.
\section{Event reconstruction and selection}
\label{sec:evSel}
\begin{figure*}[!htb]
\centering
\includegraphics[width=0.45 \textwidth]{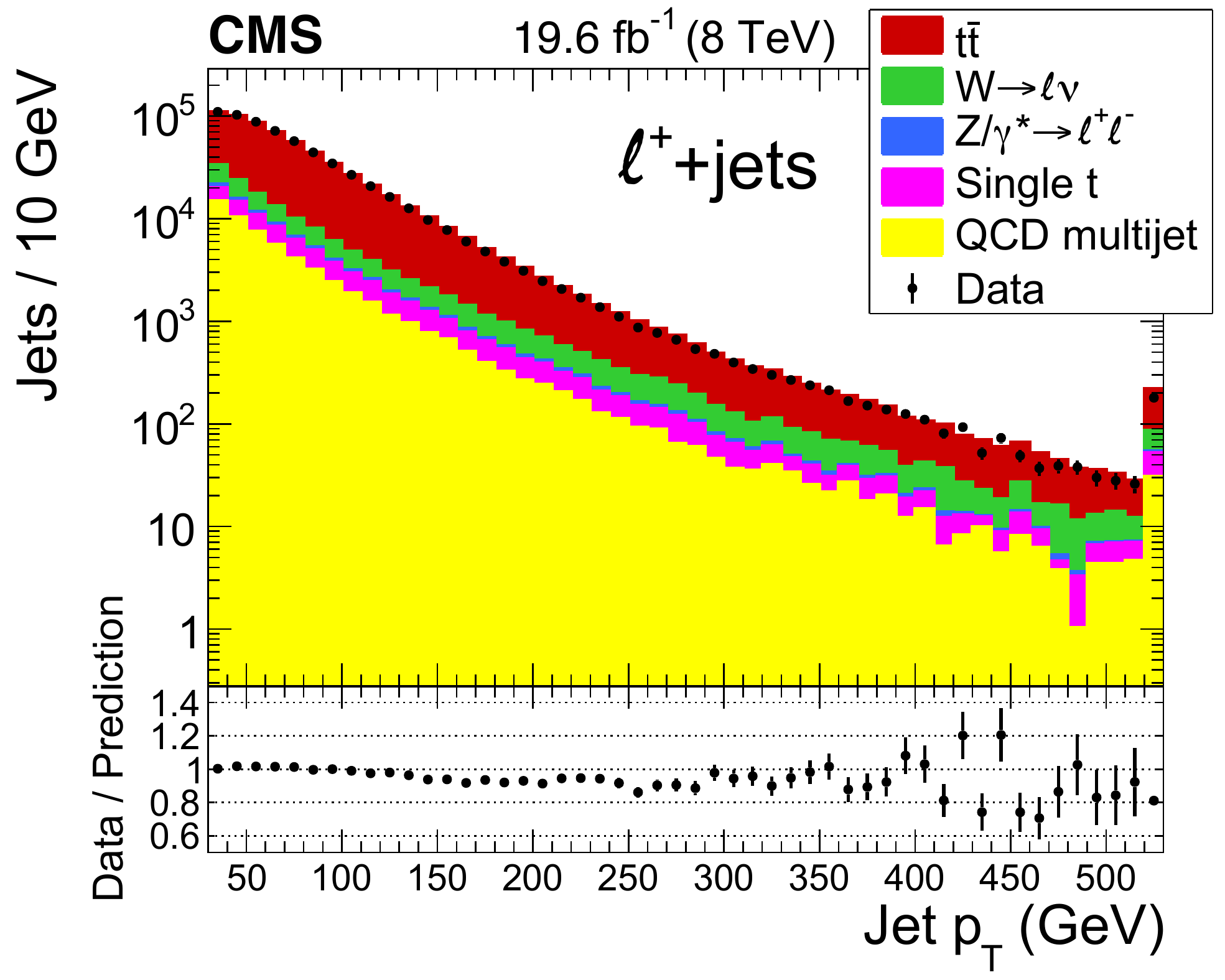}
\includegraphics[width=0.45 \textwidth]{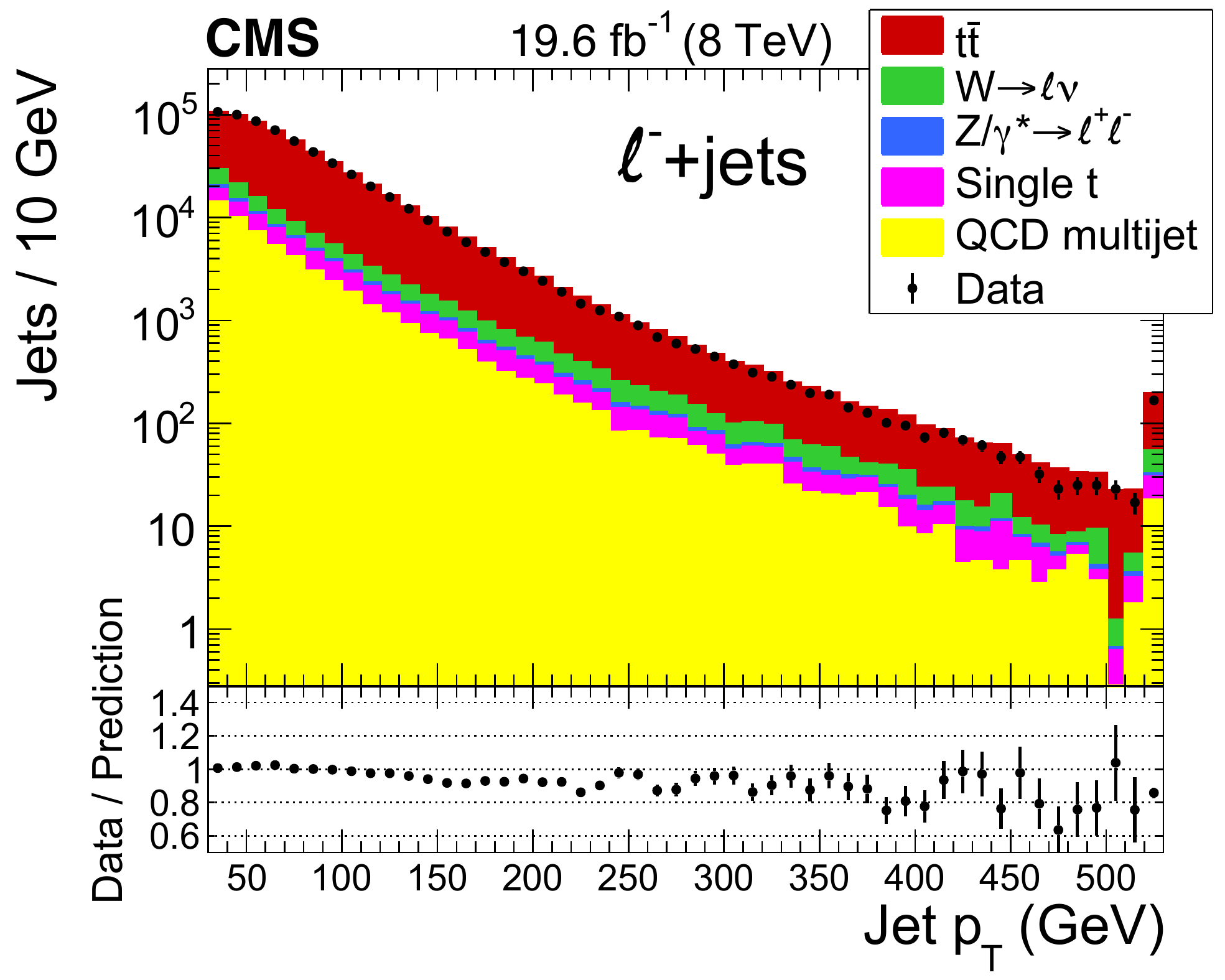}
\caption{\label{fig:dataMCselection}
Comparison of the data to expectation for the transverse momenta of the four leading jets in each event for $\ell^+$+jets events (left) and $\ell^-$+jets events (right). The last bin of each distribution includes all jets with $\pt > 530\GeV$. The bin-by-bin ratio of the observed to the simulated spectra one is shown at the bottom of each plot.
The uncertainties are purely statistical.
The total simulated event yields are normalized to the observed yields in the data, while keeping the relative fractions of the individual components fixed.
}
\end{figure*}

All events are reconstructed using the standard CMS particle-flow (PF) techniques~\cite{CMS-PAS-PFT-10-002}, where the information from all CMS subdetectors is combined in a coherent manner to identify and reconstruct individual electrons, muons, photons, charged hadrons, and neutral hadrons. Only the charged particles associated with the primary collision vertex are used in the analysis, where the primary vertex is defined as having the largest value of $\sum \pt^2$ of its associated tracks.
Reconstructed charged particles from the primary collision vertex, with the exception of isolated electrons and muons, and all neutral particles are used for jet clustering.
Jets are formed using the anti-\kt\ clustering algorithm~\cite{antikt} with a distance parameter of $0.5$. The momentum of a jet, determined from the vectorial sum of the momenta of all particles within a jet, is found from simulation to lie typically within 5--10\% of the true jet momentum. Jet energies are corrected for contributions from additional pileup interactions expected within the area of the jet.
Afterwards, simulation-based $\pt$- and $\eta$-dependent jet energy scale corrections are applied to all jets both in the data and simulation~\cite{JESpaper}.
Through these means, a uniform energy response is achieved at the reconstructed particle level with only weak pileup dependence.
Jets in the data have an additional residual correction that is determined by assuming momentum balance in dijet, photon+jet, and Z+jet events.
The jet energy resolution is measured in the data to be about 10\% worse than in simulation~\cite{JESpaper} which is corrected by smearing the jet energy in simulated events by the corresponding amount.
The amount of missing transverse momentum (\MET) is calculated as the magnitude of the vector sum of the transverse momenta of all reconstructed particles~\cite{METpaper}. The effect of jet energy scale corrections on the jet momenta is propagated to \MET.
\begin{table*}[tb]
\begin{center}
\topcaption{Expected and observed yield of events passing the full selection of $\Pep$+jets, $\Pem$+jets, $\Pgmp$+jets, and $\Pgmm$+jets channels. Simulations are used to obtain the expected number of events except for the QCD multijet background, which is derived from data, as described in the text. The uncertainties on the event numbers are statistical and reflect the limited number of events in simulation or data for the individual processes.}\label{table:eventCounts}
\begin{tabular}{lcccc}
Sample                        & $\Pep$+jets       & $\Pem$+jets       & $\Pgmp$+jets      & $\Pgmm$+jets    \\ \hline
\ttbar                            & $55922 \pm  68$ & $55476 \pm  68$ & $72020 \pm  77$ & $72094 \pm  77 $ \\
$\PW$+jets                      & $\x5448 \pm  50$ & $ \x4128 \pm  45$ & $ \x7146 \pm  59$ & $ \x5174 \pm  51 $ \\
$\Z/ \gamma^{*}$+jets & $\x\x835 \pm  12$ & $  \x\x800 \pm  11$ & $  \x\x812 \pm  12$ & $  \x\x816 \pm  11 $ \\
Single top                    &  $\x3107 \pm  35$ & $ \x2659 \pm  34$ & $ \x3908 \pm  40$ & $ \x3412 \pm  38 $ \\
QCD multijet               & $\x7922 \pm  89$ & $ \x7235 \pm  85$ & $ \x7148 \pm  85$ & $ \x7173 \pm  85 $ \\
Total                            & $\x73234 \pm 128$ & $\x70298 \pm 123$ & $\x91034 \pm 136$ & $\x88669 \pm 132 $ \\ \hline
Observed                    & $71952\, \phantom{\pm}\, \phantom{00}$ & $70396\, \phantom{\pm}\, \phantom{00}$ & $87039\, \phantom{\pm}\, \phantom{00}$ & $84024\, \phantom{\pm}\, \phantom{00}$ \\
\end{tabular}
\end{center}
\end{table*}
A particle-based relative isolation is computed for each lepton and is corrected on an event-by-event basis for contributions from pileup events~\cite{Khachatryan:2016mqs}.
The scalar sum of the transverse momenta of all reconstructed particle candidates, except for the leptons themselves, within a cone of size $\Delta R=\sqrt{\smash[b]{(\Delta \eta)^{2}+(\Delta \phi)^{2}}}=0.3$ ($=0.4$ for muons) built around the lepton direction must be less than 10\% of the electron \pt\ and less than 12\% of the muon \pt.
Events are required to contain only one isolated light lepton, either an electron with $\pt > 32\GeV$ and $|\eta| < 2.5$ or a muon with $\pt > 25\GeV$ and $|\eta|<2.1$.
Events must have at least four jets with $\pt > 30\GeV$ and $|\eta|<2.4$.
Jets originating from a bottom quark (\cPqb\ jets) are identified with the combined secondary vertex (CSV) algorithm~\cite{CMSbtvPaper,CMS-PAS-BTV-13-001} which combines the information from reconstructed secondary vertices and from displaced tracks within the jets to form a multi-variant discriminator output.
It is tuned such that its efficiency to tag \cPqb\ jets is about $68\%$ and the rate of mistagging light-flavor, gluon, and \cPqc\ jets is about $4\%$ for jets within the considered $\pt$ range, as evaluated using the nominal \ttbar simulation.
Each event is required to have at least one \cPqb-tagged jet. An additional event selection requirement based on the $\chi^2$ of a kinematic fit to the \ttbar\ hypothesis described in Section~\ref{sec:kinfit} is also applied in the analysis.
The number of events observed in the data and the corresponding predictions from simulated events and from data control regions (for the QCD multijet background) are shown in Table~\ref{table:eventCounts} separately for $\Pep$+jets, $\Pem$+jets, $\Pgmp$+jets, and $\Pgmm$+jets events.
The fraction of QCD multijet events is estimated with a binned maximum-likelihood fit to the \MET\ distribution observed in the data, separately in the $\Pe$+jets and $\Pgm$+jets channels. The mass shape for the QCD multijet background is obtained from the data by using a dedicated sample of events where the isolation and identification criteria (for $\Pe$+jets) or the isolation criterion alone (for $\Pgm$+jets) have been inverted.
A difference of less than 6\% is found between the observed and expected total yields.
Differences between the data and expectation in the overall yield do not affect this analysis directly, unlike possible differences in the kinematic properties of the events or in the relative fractions of the yields of the various processes.
Figure~\ref{fig:dataMCselection} shows the distribution of the transverse momenta of the four leading jets in each event for both the $\ell^+$+jets and the $\ell^-$+jets samples.
The overall number of simulated events is normalized to the event yield observed in the data, while keeping their relative fractions fixed to the prediction.
In general, the data appear to be well modeled by the simulation, except for a small but statistically significant deviation in the jet transverse momenta, visible as a slope in the ratio plots.
This is related to the modeling of the top quark transverse momentum in the simulation, which is known to predict a slightly harder spectrum than observed in data~\cite{Chatrchyan:2012saa,Khachatryan:2015oqa,Khachatryan:2015fwh}.
The effect on the top quark mass measurement is approximately 100\MeV, as measured in an equivalent analysis in the $\ell$+jets channel~\cite{CMSTopMassLegacy}.
As it identically affects the $\ell^+$+jets and $\ell^-$+jets samples, the impact on the $\Delta m_{\cPqt}$ measurement is negligible.
\section{The kinematic fit and the ideogram method}
\label{sec:kinfit}
\begin{figure*}[!htb]
\centering
\includegraphics[width=0.45 \textwidth]{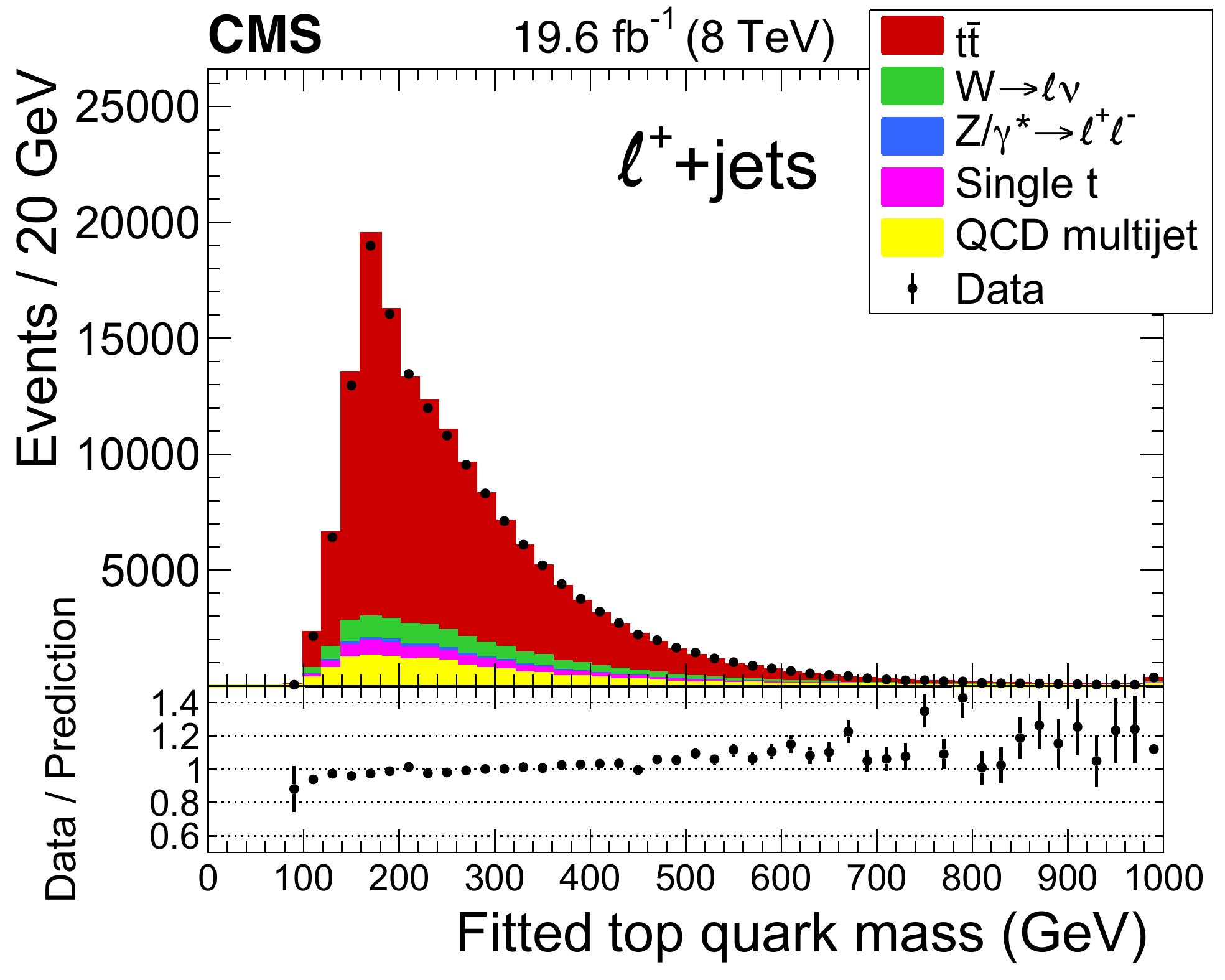} \hspace{0.5cm}
\includegraphics[width=0.45 \textwidth]{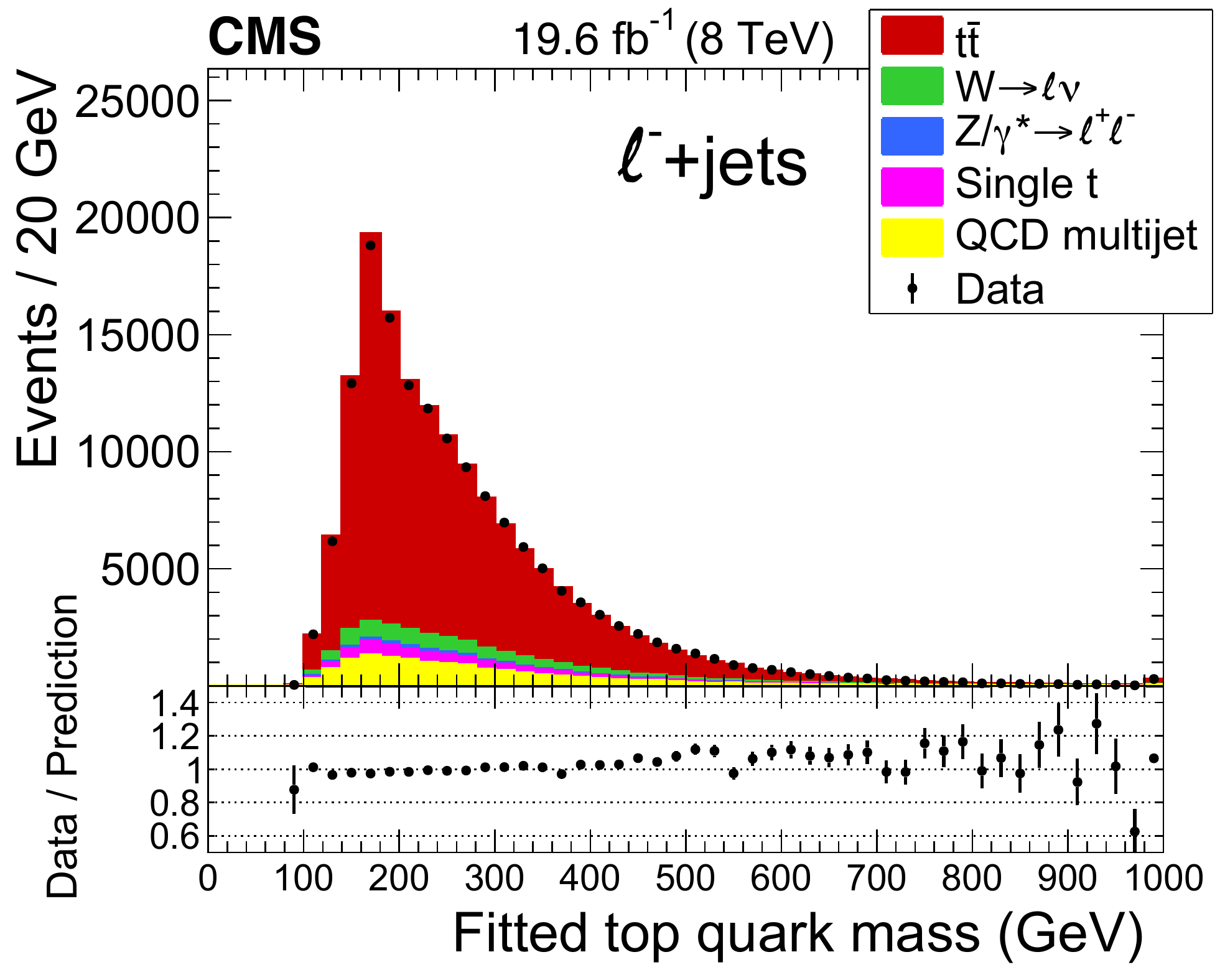} \\ \vspace{0.3cm}
\includegraphics[width=0.45 \textwidth]{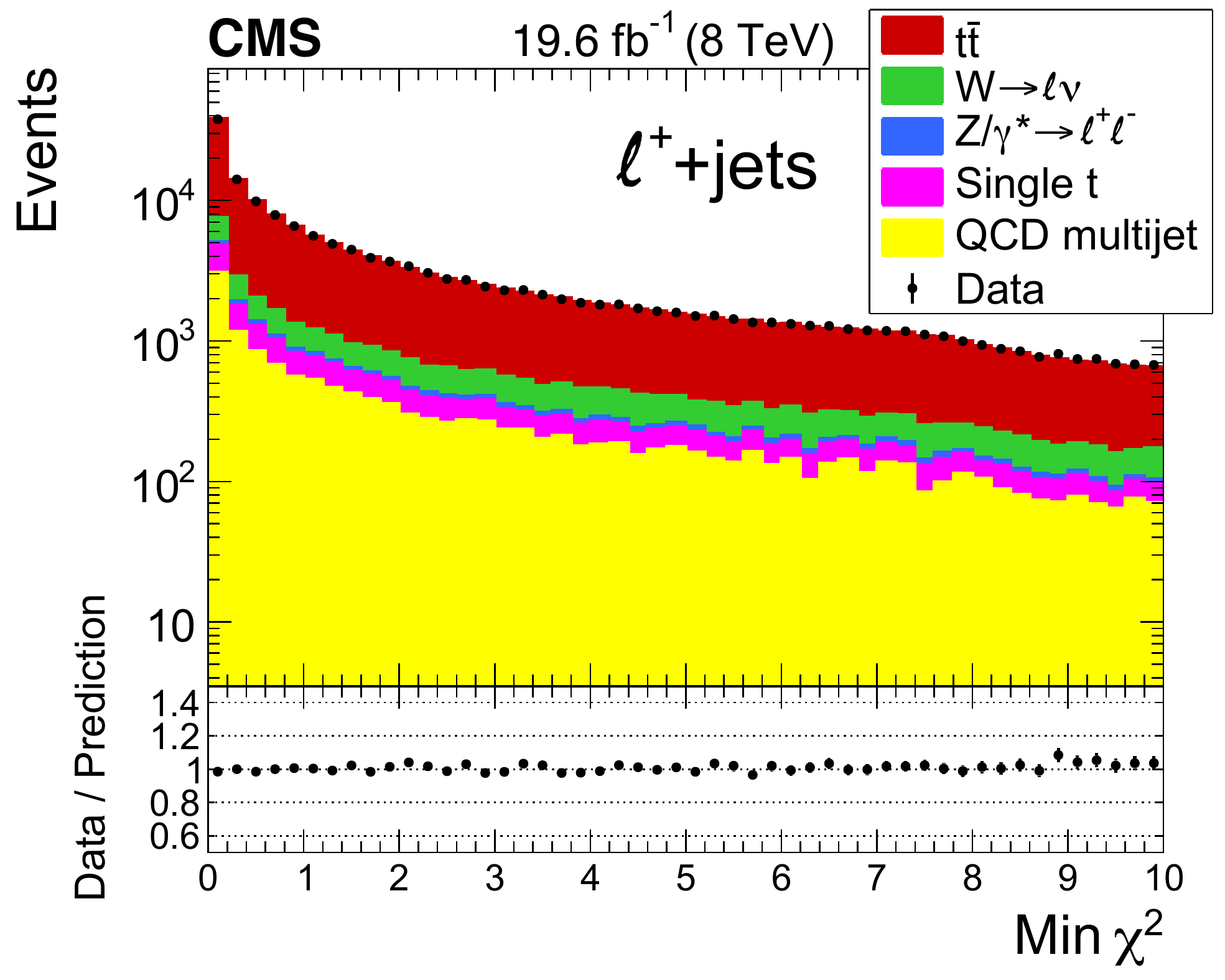} \hspace{0.5cm}
\includegraphics[width=0.45 \textwidth]{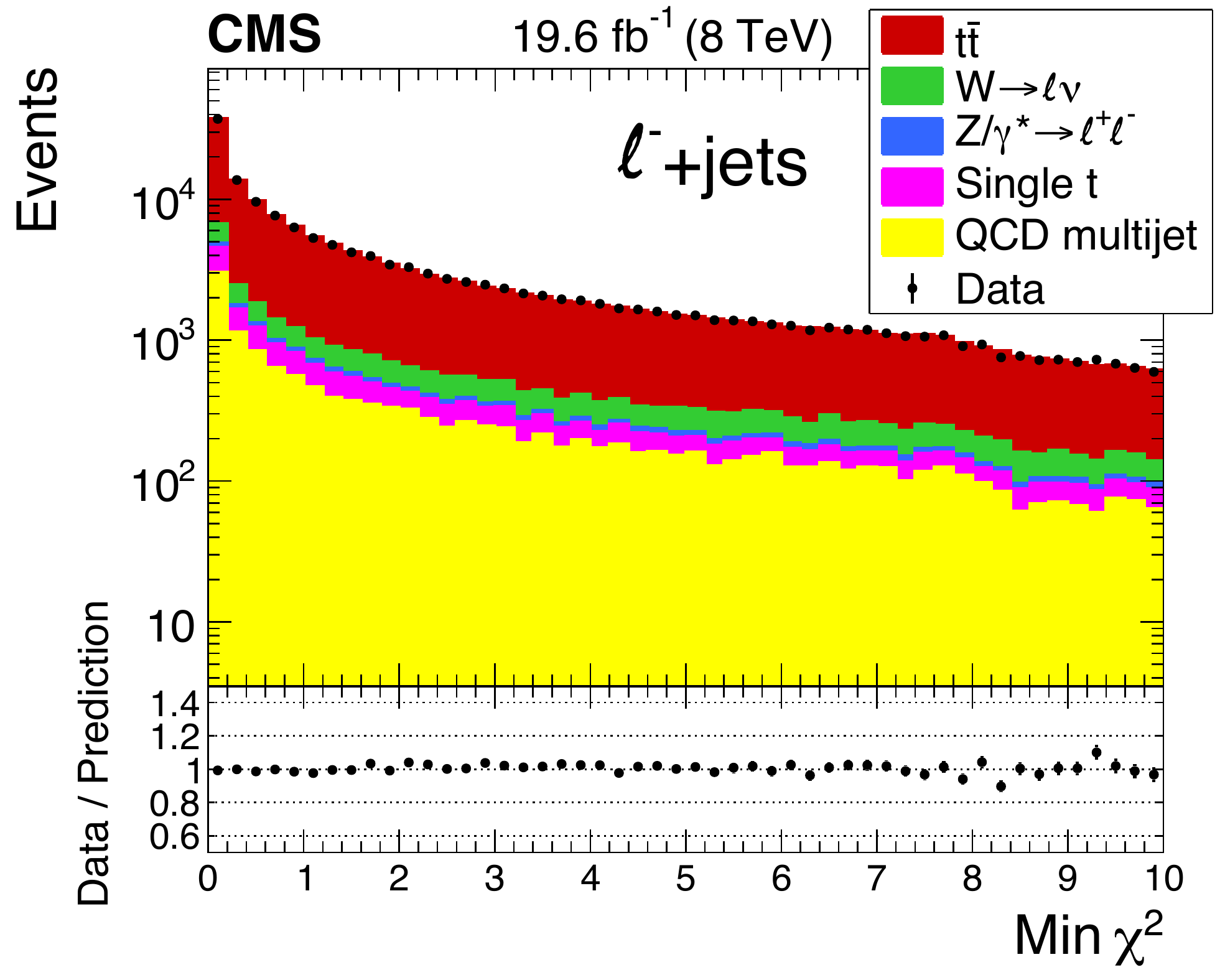}
\caption{
Comparison between the data and expectation for the fitted top quark mass of the jet-quark assignment with the smallest $\chi^2$ (top) and these smallest $\chi^2$ values (bottom), for $\ell^+$+jets events (left) and $\ell^-$+jets events (right). The last bin of the top quark mass distributions includes all masses above 980\GeV. The bin-by-bin ratio of the observed spectrum to the simulated one is shown at the bottom of each plot. The uncertainties are purely statistical.
\label{fig:dataMCkinfit}
}
\end{figure*}

\begin{figure*}[tb]
\centering
\includegraphics[width=0.49 \textwidth]{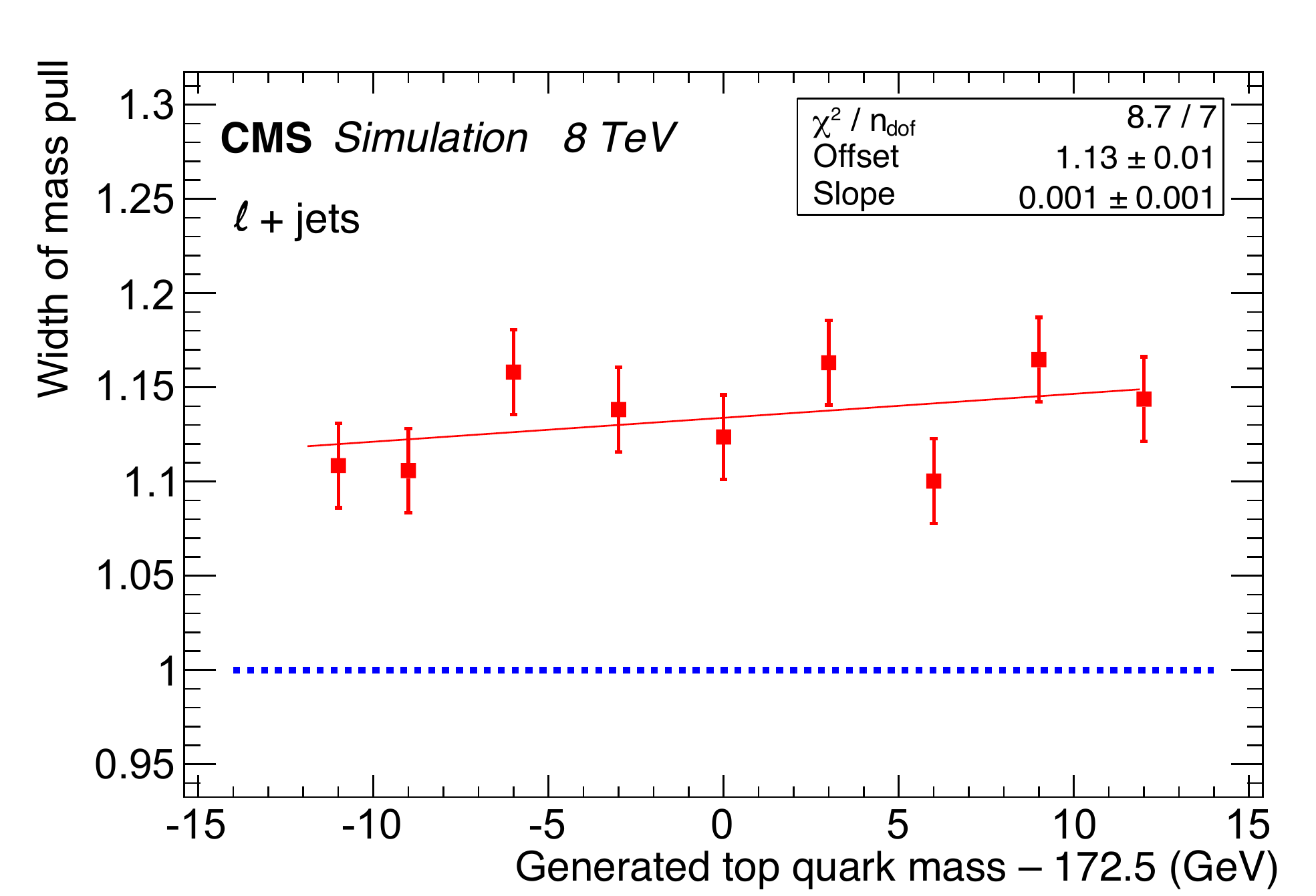}
\includegraphics[width=0.49 \textwidth]{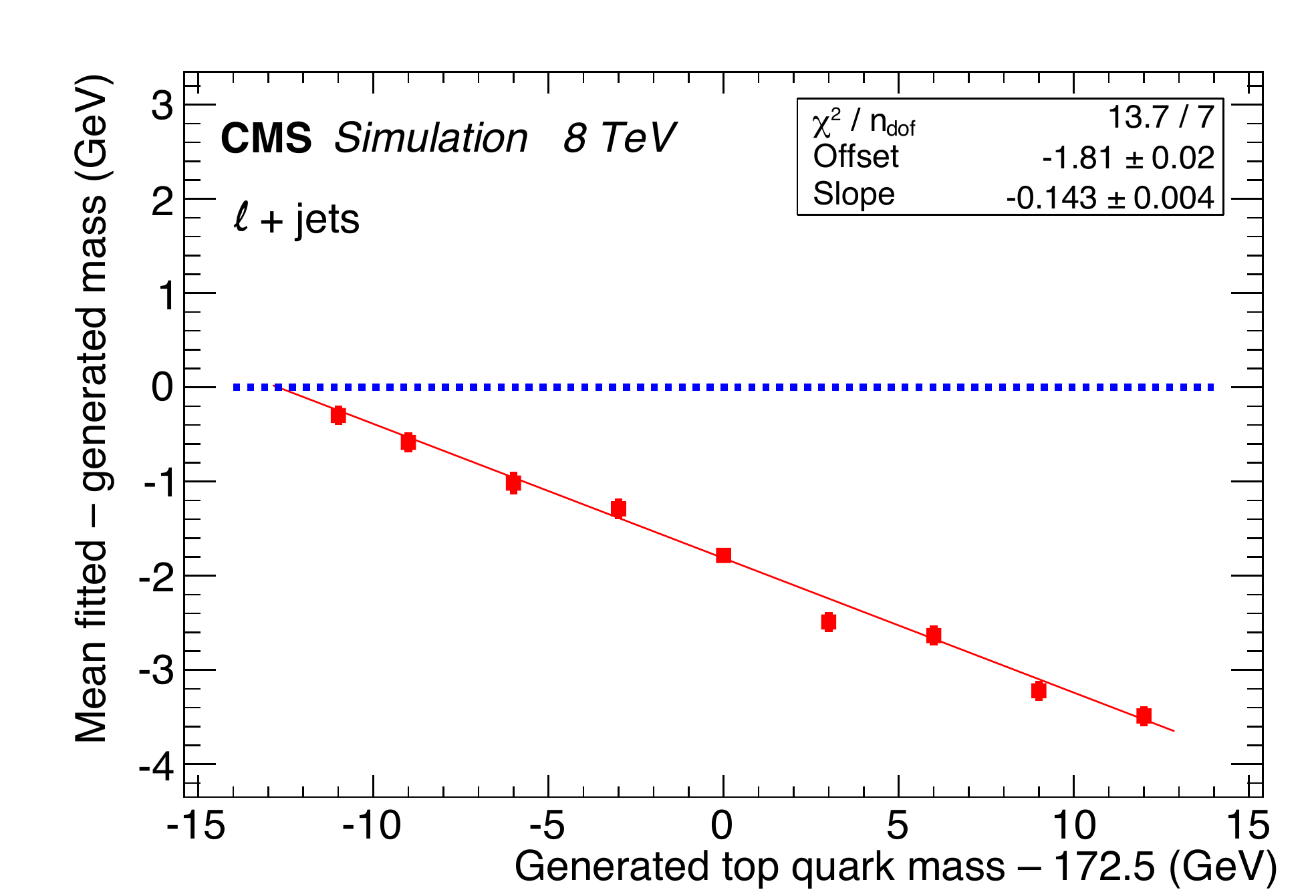}
\caption{Width of the pull distribution (left) and bias on the estimated top quark mass (right) as a function of the generated top quark mass for $\ell$+jets events. The dashed blue line represents the ideal outcome.\label{fig:calibPullBias}}
\end{figure*}

\begin{figure*}[tb]
\centering
\includegraphics[width=0.49 \textwidth]{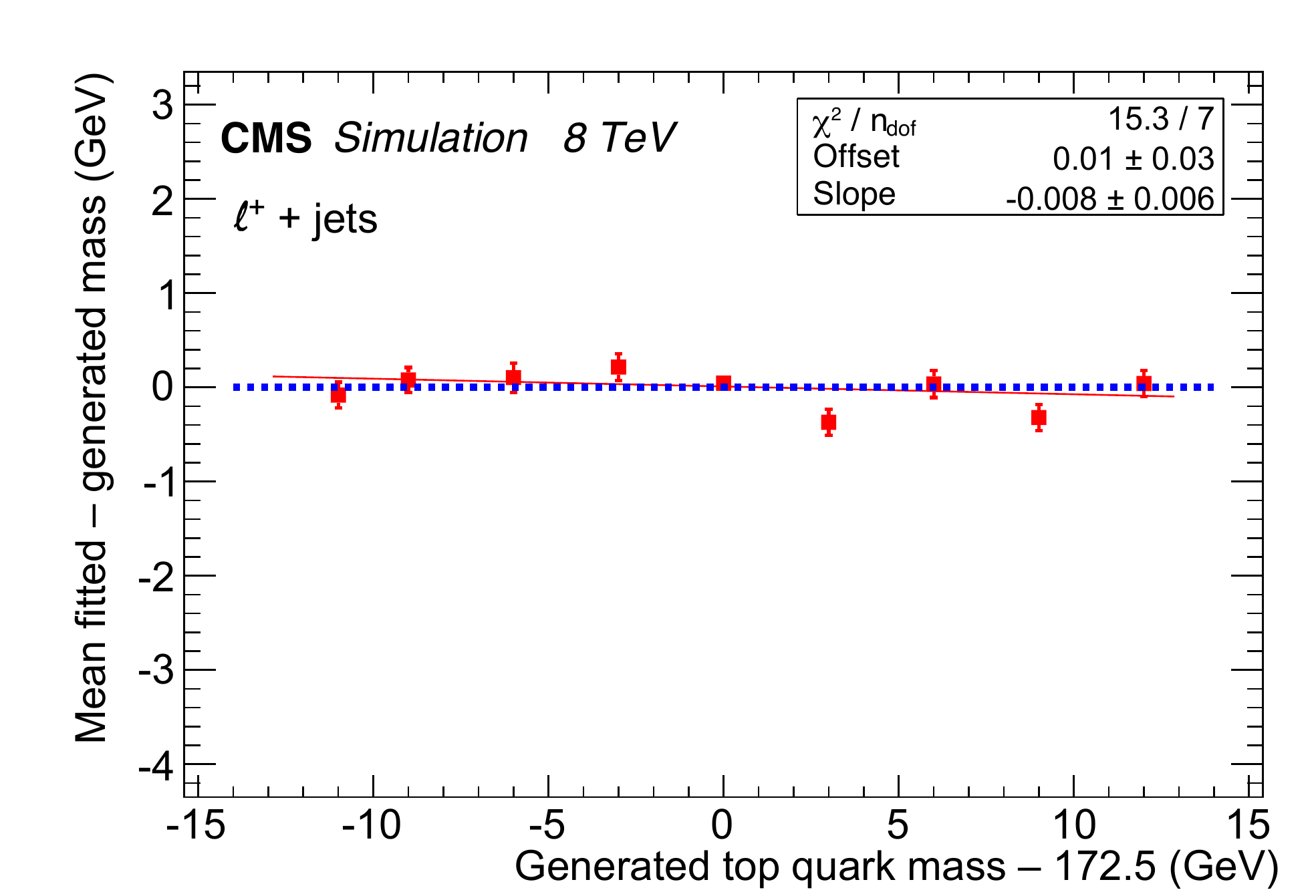}
\includegraphics[width=0.49 \textwidth]{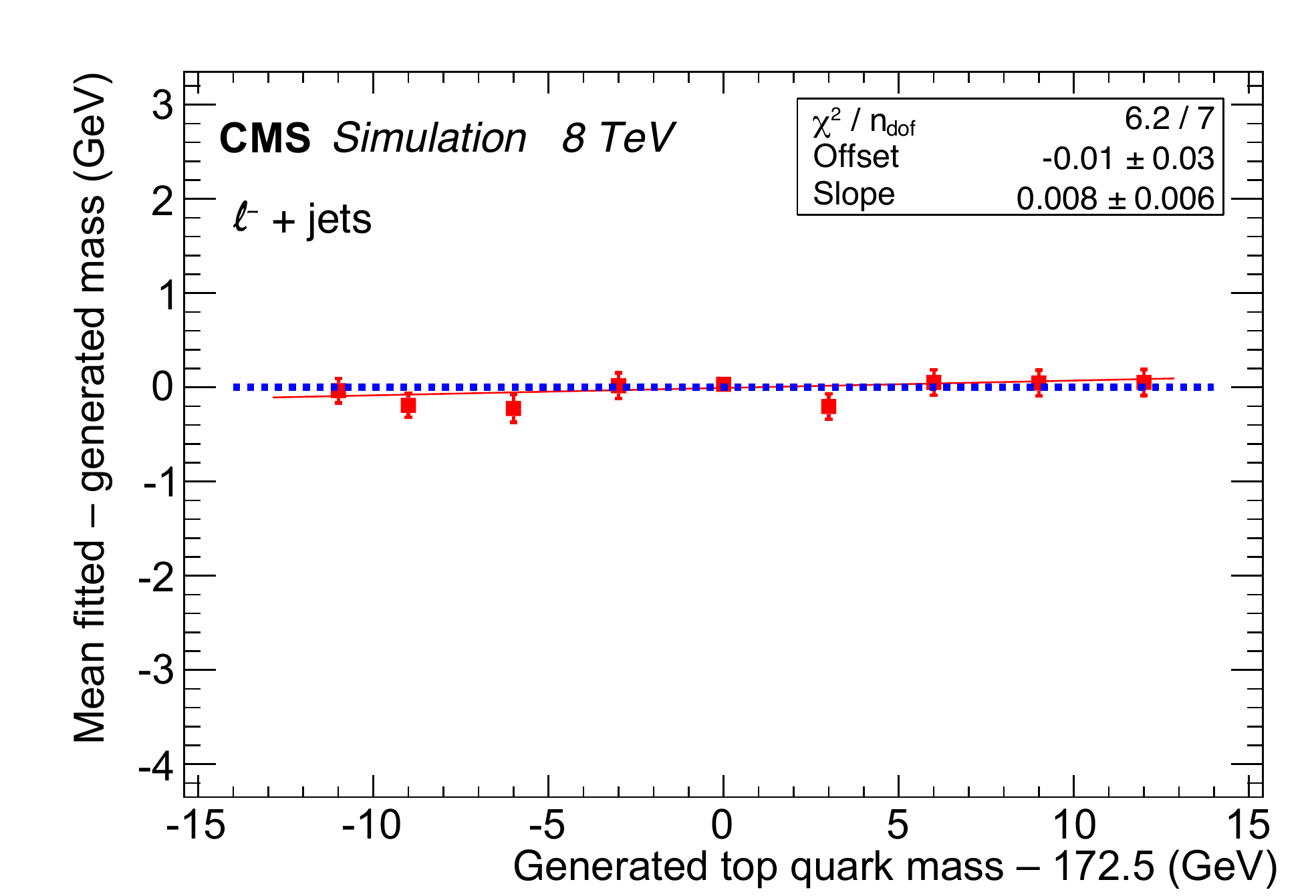}
\caption{Residual bias on the estimated top quark mass as a function of the generated top quark mass using $\ell^+$+jets events (left) and $\ell^-$+jets events (right) after the inclusive $\ell$+jets calibration. The dashed blue line represents the ideal outcome.\label{fig:calibratedBias}}
\end{figure*}

A kinematic fit to the \ttbar\ hypothesis~\cite{ThesisPetra,KinFitNote,Blyweert:1967363} is employed in this analysis to reconstruct the mass of the hadronically decaying top quark by varying the momenta of the two jets that are assigned to the $\PW \to \cPq\cPaq'$ decay, using a \PW\ boson mass value of $80.4\GeV$~\cite{PDG} as a constraint, while keeping the ratio of energy to momentum of each jet fixed.
The $\PWp$ and $\PWm$ bosons are assumed to have equal mass in this procedure.
For each event, the four jets with the largest transverse momentum are considered in the fit.
These four jets can be associated with the four quarks from the hypothesized \ttbar-decay ($\ttbar \rightarrow \text{b}\overline{\text{b}}\PWp\PWm \rightarrow \text{b}\overline{\text{b}} \text{q} \overline{\text{q}}' \ell\nu_{\ell}$) in 12 possible ways.
The kinematic fit is performed for each of these 12 jet-to-quark assignments.
However, before carrying out the fit, additional corrections are applied to the data and simulation in order to correct jet energies to the parton level.
These corrections are derived separately for jets associated with hadronic \PW\ boson decays and for \cPqb\ jets in bins of \pt\ and $|\eta|$ of these jets by comparing the transverse energy of reconstructed jets with that of the corresponding generated partons in simulated \ttbar\ events.
Only solutions for which the kinematic fit returns a $\chi^2/n_{\rm dof} <10$ are accepted, where $n_{\rm dof}$ ($= 1$) is the number of degrees of freedom in the fit.
An event is rejected if no combination of jets passes the $\chi^2$ requirement.
For each combination of jets $i$, $m_i$ is the top quark mass value that yields the smallest $\chi^2_{i}$, and the uncertainty $\sigma_i$ corresponds to the mass range that is compatible with an increase of the $\chi^2$ by $1$.
The values of $m_i$, $\sigma_i$, and $\chi^2_{i}$ are used as input to the ideogram method as described below.
A comparison of $m_i$ and $\chi^2_{i}$ between the data and expectation is given in Fig.~\ref{fig:dataMCkinfit}, for the jet combination with the smallest overall $\chi^2$ in each event. Agreement is observed between the data and expectation.
In the ideogram method~\cite{Abdallah:2008xh}, an approximate event-by-event likelihood model is constructed as a function of $m_{\cPqt}$. This likelihood contains a signal and a background term.
The signal part is a weighted sum over all combinations of jets, containing two terms: a correct jet-to-quark assignment term and a wrong jet-to-quark assignment term.
The shapes of the background term and the wrong jet-to-quark assignment term are both taken from simulation, which represents our best knowledge of the kinematic properties of the events.
The correct jet-to-quark assignment term, on the other hand, is defined by the convolution of a Gaussian resolution function and a relativistic Breit-Wigner distribution.
The Gaussian function has a width equal to the uncertainty $\sigma_i$ resulting from the kinematic fit for the given combination of jets.
The weights applied in the sum over jet combinations are calculated for each combination using the $\chi^2_i$ of the kinematic fit and the compatibility of the \cPqb\ jet assignments, calculated from the known \cPqb\ tagging efficiency and mistagging rate.
The reweighting significantly reduces the contribution of combinations for which the \cPqb\ jet assignments are highly incompatible with the results of the \cPqb{} tagging algorithm and of combinations that badly fulfill the mass constraints.
The combined likelihood for the full event sample is calculated as the product of the individual event likelihoods for all selected events.
The fitted top quark mass and its statistical uncertainty are extracted from this combined likelihood.
More details about the exact implementation of the kinematic fit and the ideogram method can be found in Ref.~\cite{CMStopMassDiff}.
The event likelihoods are calculated under certain assumptions and simplifications and hence the result of the combination is first calibrated using pseudo-experiments where the mass shapes are generated from the expected distributions of signal and background events.
The standard deviation (width) of the pull distribution and the observed bias on the estimated top quark mass before calibration are shown as a function of the generated mass in Fig.~\ref{fig:calibPullBias}. The pull is defined as $\text{pull}_j = (m_j - \langle m \rangle ) / \sigma_j$, where $m_j$ is the estimated top quark mass in each pseudo-experiment $j$, $\sigma_j$ the corresponding statistical uncertainty, and $\langle m \rangle$ the mean of the estimated top quark masses over all pseudo-experiments.
Since the width of the pull distribution is about 1.13, the statistical uncertainty of the final mass measurement needs to be scaled up by about 13\%.
The biases are within 3\GeV\ for most of the range of interest. The obtained top quark mass is corrected using the fitted linear function shown in Fig.~\ref{fig:calibPullBias}~(right).
The residual bias on the estimated top quark mass as a function of the generated top quark mass after applying this calibration is shown in Fig.~\ref{fig:calibratedBias}, separately for $\ell^+$+jets and $\ell^-$+jets events.
These plots demonstrate that an inclusive $\ell$+jets calibration can be used for both the positive and negative channels.
\section{Measurement of \texorpdfstring{$\Delta m_{\cPqt}$}{Delta m[t]}}
\label{sec:diffMeasurement}
The analysis is applied separately to events with positively and negatively charged leptons, for each of which the top quark mass is measured using the hadronically decaying top quark.
The difference between the two resulting mass measurements is then taken as the final measurement of the mass difference between the top quark and antiquark.
In the inclusive $\Pe$+jets and $\Pgm$+jets sample a mass difference of $\Delta m_{\cPqt} = -0.15 \pm 0.19\stat\GeV$ is measured.
In the individual $\Pe$+jets and $\Pgm$+jets channels, respective mass differences of $\Delta m_{\cPqt} = -0.19 \pm 0.28\stat\GeV$ and $\Delta m_{\cPqt} = -0.13 \pm 0.26\stat\GeV$ are obtained.
These results are compatible with the hypothesis of CPT conservation. The average top quark mass is measured to be $m_{\cPqt} = 172.84 \pm 0.10\stat\GeV$, which is in agreement with previous measurements~\cite{CMSTopMassLegacy,TopMassWorldAv, Aaboud:2016igd}, even ignoring systematic uncertainties.
\section{Systematic uncertainties}
\label{sec:systematics}
Many of the systematic uncertainties that affect the top quark mass measurement have a significantly reduced impact in the mass difference because of their correlated effect on the individual top quark and antiquark mass extractions.
Some systematic uncertainties related to the modeling of the physics processes are not expected to affect the $\Delta m_{\cPqt}$ measurement and are not considered in this analysis.
These are the modeling of the hadronization, the underlying event, initial- and final-state radiation, the factorization and renormalization scales, and the matching between matrix-element and parton shower calculations.
Other effects considered in the measurement of $m_{\cPqt}$ are included together with additional sources potentially relevant for the $\Delta m_{\cPqt}$ measurement, such as lepton-charge identification and a possible difference in jet energy response to \cPqb\ and \cPaqb\ quarks.
A summary of these effects is given in Table~\ref{table:Systematics}.
The effects are evaluated by comparing the nominal simulation to a sample of simulated events where the source of the systematic uncertainty under study is varied within its uncertainty.
Since most sources of systematic uncertainty yield only a small change in the $\Delta m_{\cPqt}$ measurement, statistical uncertainties on the observed changes are evaluated using a jackknife re-sampling technique~\cite{Jackknife} and the larger among the estimated change and its statistical uncertainty is quoted as the final systematic uncertainty. The total systematic uncertainty is taken to be the quadratic sum of all individual values.
The uncertainties presented here are significantly smaller than those reported in Ref.~\cite{CMStopMassDiff}.
All systematic uncertainties in the previous result were statistically compatible with zero and the total uncertainty included a sizable component from the limited size of the simulated data samples. Much larger samples of simulated signal and background events have been produced for this new result, resulting in more accurate estimates of the systematic uncertainties and a reduction of the total uncertainties.
Some uncertainties, such as the jet energy scale and the \cPqb\ tagging efficiency also profit from more accurate corrections.
\begin{table}[htb]
\begin{center}
\topcaption{Summary of systematic uncertainties on $\Delta m_{\cPqt}$. For each contribution, the first value is the observed systematic shift, whereas the second number is the uncertainty of the shift due to the limited number of generated events. In all cases, the larger among the two is considered as the final systematic uncertainty and is indicated in the bold font. The total uncertainty is obtained from the sum in quadrature of the individual terms.\label{table:Systematics}}
\begin{tabular}{lc}
Source                                & \begin{tabular}{c}Uncertainty \\ in $\Delta m_{\cPqt}$  (\MeVns)  \end{tabular}      \\ \hline
Jet energy scale                      & $       \x7  \pm {\bf 16} $ \\
Jet energy resolution                 & $       \x7  \pm {\bf 11} $ \\
\cPqb\ vs. \cPaqb\ jet response       & $ {\bf 51} \pm       1\x  $ \\
Signal fraction                       & $ {\bf 27} \pm       2\x  $ \\
Background charge asymmetry           & $ {\bf 11.9} \pm  0.1\x  $ \\
Background composition                & $ {\bf 28} \pm       1\x  $ \\
Pileup                                & $ {\bf  9.1} \pm       0.3  $ \\
\cPqb\ tagging efficiency             & $ {\bf 24} \pm       7\x  $ \\
\cPqb\ vs. \cPaqb\ tagging efficiency & $ {\bf 11} \pm       7\x  $ \\
Method calibration                    & $       \x3  \pm {\bf 53} $ \\
Parton distribution functions         & $ {\bf  9} \pm       3  $ \\ \hline
Total                                 & { 91 }\\
\end{tabular}
\end{center}
\end{table}

The individual contributions to the total systematic uncertainty are described in more detail below.
\begin{description}
\item[Jet energy scale.] Since top quarks and antiquarks are produced at the LHC with slightly different rapidity distributions, the $\eta$-dependence of the jet energy scale uncertainty can lead to an effect on $\Delta m_{\cPqt}$. To evaluate this effect the energy of all jets is scaled up/down within their $\PT$- and $\eta$-dependent uncertainties (ranging between 1 and 5\%)~\cite{JESpaper}. This results in a shift in mass difference of $7 \pm 16\MeV$.
\item[Jet energy resolution.] To evaluate the systematic uncertainty arising from the uncertainty in the measured jet energy resolution, the jet energy in simulation is smeared up/down within the uncertainty of this jet energy resolution.
The jet energy resolution uncertainty is $|\eta|$-dependent and ranges between 6 and 9\% for the jets considered in this analysis.
A shift of $7 \pm 11\MeV$ in $\Delta m_{\cPqt}$ is observed.
\item[\cPqb\ vs. \cPaqb\ jet response.] A difference in the fraction of jet energy reconstructed by the detector between \cPqb\ and \cPaqb\ jets can introduce a bias in the $\Delta m_{\cPqt}$ measurement. Such differences, caused for example by different cross sections for interactions of positively and negatively charged kaons in the calorimeter, are expected to be reduced thanks to the PF reconstruction that relies mostly on tracking to reconstruct charged hadrons. The $\pt$ of the reconstructed jets is compared with the original parton $\pt$ in two simulated \ttbar\ samples: the nominal sample, generated with \MADGRAPH\ with showering from \PYTHIA, and a sample generated with \textsc{mc@nlo} with showering from \HERWIG.\@ Simulated samples produced with these two sets of MC generators have been observed to encompass the data in various key observables, and differ significantly in several aspects, including the relative production and decay rates of different kinds of hadrons in the jets~\cite{ATLASGeneratorComparison}. In both samples the ratio of \cPqb\ to \cPaqb\ response as a function of jet $\PT$ is observed to be statistically compatible with unity, and an average difference of $0.078 \pm 0.040$\% is measured. When the difference of 0.078\% is propagated to our nominal sample of simulated events a shift of $51 \pm 1\MeV$\ is observed, which is quoted as systematic uncertainty.
\item[Signal fraction.] A change in the signal fraction (as calculated from Table~\ref{table:eventCounts}) will bias the measured top quark mass, since signal and background events have different fitted top quark mass distributions. This will also introduce a bias in $\Delta m_{\cPqt}$ because it will influence the $\ell^+$+jets and $\ell^-$+jets samples in a different way since these have a different signal fraction. The signal fraction is changed by a relative ${\pm} 10$\%, corresponding to the agreement between the expected and observed \ttbar\ cross sections in this channel~\cite{Khachatryan:2016yzq}, and the resulting shift of $27 \pm 2\MeV$\ is taken as a systematic uncertainty.
\item[Background charge asymmetry.] A difference in the estimated charge asymmetry of the backgrounds leads to different levels of background and to a different background composition in the $\ell^+$+jets and $\ell^-$+jets channels, which can bias the $\Delta m_{\cPqt}$ measurement. The measured inclusive $\PWp/\PWm$ production ratio at 8\TeV is in agreement with theoretical predictions within a precision of 2\%~\cite{ChargeAsymmetry,Chatrchyan:2014mua}, but since this ratio depends on the number of jets, the uncertainty is inflated by a factor of two, yielding a variation of 4\%. When the fractions of $\PWp$ and $\PWm$ events are varied by 2\% in opposite directions, thereby affecting the relative ratio of $\PWp$ and $\PWm$ events by 4\%, $\Delta m_{\cPqt}$ changes by $3.72 \pm 0.01\MeV$. The $\PW$+jets background contains non-negligible contributions from $\PW$+$\cPqc\cPaqc$ and $\PW$+$\cPqb\cPaqb$ events, whose relative $\PWp$/$\PWm$ ratio is affected by a larger uncertainty. The relative ratio is varied by 20\%, which corresponds to the uncertainty in the measured inclusive $\PW$+$\cPqb\cPaqb$ production cross section~\cite{CMSWbb}, and yields a shift in $\Delta m_{\cPqt}$ of $9.05 \pm 0.02\MeV$ and $5.83 \pm 0.02\MeV$ for the $\PWp\cPqc\cPaqc$ and $\PWp\cPqb\cPaqb$ contributions, respectively. Single top quarks produced via the $t$ channel also possess a charge asymmetry, measured to be in agreement with theory predictions within 15\%~\cite{ChargeAsymmetrySingleTop}.
Changing this charge asymmetry by a relative ${\pm}15\%$ results in a shift on $\Delta m_{\cPqt}$ of $3.298 \pm 0.005\MeV$. The quadratic sum of all these observed shifts is quoted as the systematic uncertainty.
\item[Background composition.] Possible residual effects due to the composition of the background are evaluated by scaling each background source up and down, keeping the signal fraction fixed. A shift in $\Delta m_{\cPqt}$ is observed when we scale $\PW$+jets ($1.3 \pm 0.3\MeV$), $\PZ$+jets ($1.99 \pm 0.03\MeV$), $t$-channel single top quark production ($6.9 \pm 0.1\MeV$), and $\cPqt\PW$ single top quark production ($1.4 \pm 0.3\MeV$) up/down by 30\%; and when we scale QCD multijet events ($26.8 \pm 0.3\MeV$) up/down by 50\%. The size of each variation was chosen to cover the modeling uncertainty in predictions of the MC simulation in the phase space of the analysis or, in the case of the QCD multijet sample, differences between estimates obtained with different methods to determine the normalization from data.
The systematic uncertainty is obtained by summing in quadrature each of the observed shifts.
\item[Pileup.] Pileup collisions are included in the sample of simulated events used in this analysis. Events are reweighted to reproduce the pileup distribution measured in the data. The systematic uncertainty is estimated by changing the mean value of the number of interactions by $\pm 6\%$ to account for uncertainties in the rate~\cite{Lumi} and exact properties of the pileup collisions. This results in a shift in $\Delta m_{\cPqt}$ of $9.1 \pm 0.3\MeV$.
\sloppy{
\item[$\cPqb$ tagging efficiency and $\cPqb$ vs. $\cPaqb$ tagging efficiency.] A mismodeling in simulation of the $\cPqb$ tagging efficiency can bias the measurement by altering the observed $\cPqb$ tagging assignments, which are used in the ideogram method. To quantify the impact of the uncertainty in the $\cPqb$ tagging efficiency, we change the working point of the $\cPqb$ tagging algorithm. Working points corresponding to an absolute change of ${\pm}1.2$\%~\cite{CMSbtvPaper,CMS-PAS-BTV-13-001} in the $\cPqb$ tagging efficiency produce a shift in $\Delta m_{\cPqt}$ of $24 \pm 7\MeV$\ (``$\cPqb$ tagging efficiency'' in Table~\ref{table:Systematics}). The use of different working points for the $\ell^+$+jets and $\ell^-$+jets samples, yielding an absolute 1.2\% difference in $\cPqb$ tagging efficiency between $\cPqb$ and $\cPaqb$ jets, produces a shift of $11 \pm 7\MeV$ (``$\cPqb$ versus $\cPaqb$ tagging efficiency'' in Table~\ref{table:Systematics}).}
\item[Misassignment of lepton charge.] In this analysis the leptons are only used in the trigger, in the event selection, and in the splitting of the data into $\ell^+$+jets and $\ell^-$+jets samples, but not in the mass reconstruction. A misassignment of the lepton charge can affect the calibration and it can also lead to a dilution of the measurement. For muons the charge misassignment rate is measured with cosmic muons~\cite{MuonPerformance} and collision data~\cite{ChargeAsymmetry,Chatrchyan:2014mua} to be of the order of $10^{-5}$ to $10^{-4}$ in the considered $\pt$ range. For electrons this rate ranges between 0.1\% and 0.4\%~\cite{ChargeAsymmetry,Chatrchyan:2014mua}. This means that the systematic uncertainty from charge misassignment is below 1\% of the measured $\Delta m_{\cPqt}$ value, which is negligible and is therefore ignored.
\item[Trigger, lepton identification, and lepton isolation.] As the trigger is based on an isolated single lepton, and the lepton is not used in the mass reconstruction, no systematic effect is expected from an uncertainty in the trigger efficiency or on the lepton energy scale. Similarly, the lepton identification and isolation are also not expected to affect the measurement.
\item[Method calibration.] The difference in mass between the $\ell^+$+jets and $\ell^-$+jets samples in the nominal \textsc{MadGraph}+\PYTHIA\ sample with $m_{\text t} = 172.5\GeV$, is found to be $3 \pm 53\MeV$. This result is statistically compatible with zero and confirms our expectation that there is no known effect in simulation that would lead to a difference in mass calibration between the two channels. The statistical uncertainty is quoted as the systematic uncertainty arising from the method calibration.
As a further cross-check, events are reweighted to simulate a difference in mass between top quarks and antiquarks in the nominal sample, ranging in small steps from $-4$ to $+4\GeV$. A linear relation between simulated and measured mass difference is observed, with a slope compatible with unity, and a statistical precision of 5\%. If propagated to the final result, this uncertainty in slope would have a negligible impact on the final uncertainty.
\item[Parton distribution functions.] The choice of the parton distribution functions (PDFs) can affect the $\Delta m_{\cPqt}$ measurement in multiple ways. They determine, for example, the difference in production of $\PWp$ and $\PWm$ events. The simulated samples are generated using the CTEQ 6.6 PDF~\cite{CTEQ}, for which the uncertainties can be described by 22 independent parameters. Varying each of these parameters within the quoted uncertainties and summing the larger shifts in quadrature results in an uncertainty in $\Delta m_{\cPqt}$ of $9 \pm 3\MeV$.
\end{description}

\ifthenelse{\boolean{cms@external}}{}{\clearpage}

\section{Results and summary}\label{sec:conclusion}
Data collected by the CMS experiment in pp collisions at $\sqrt{s} = 8\TeV$ and corresponding to an integrated luminosity of $19.6 \pm 0.5 \fbinv$ have been used to measure the difference in mass between the top quark and antiquark. The measured value is
\begin{equation*}
\Delta m_{\cPqt} = -0.15 \pm 0.19\stat \pm 0.09\syst\GeV.
\end{equation*}
This result improves in precision upon previously reported measurements~\cite{CMStopMassDiff,CDFmassDiff,D0massDiff,Aaltonen:2012zb,ATLASmassDiff} by more than a factor of two.
It is in agreement with the expectations from CPT invariance, requiring equal particle and antiparticle masses.

\begin{acknowledgments}

We congratulate our colleagues in the CERN accelerator departments for the excellent performance of the LHC and thank the technical and administrative staffs at CERN and at other CMS institutes for their contributions to the success of the CMS effort. In addition, we gratefully acknowledge the computing centres and personnel of the Worldwide LHC Computing Grid for delivering so effectively the computing infrastructure essential to our analyses. Finally, we acknowledge the enduring support for the construction and operation of the LHC and the CMS detector provided by the following funding agencies: BMWFW and FWF (Austria); FNRS and FWO (Belgium); CNPq, CAPES, FAPERJ, and FAPESP (Brazil); MES (Bulgaria); CERN; CAS, MoST, and NSFC (China); COLCIENCIAS (Colombia); MSES and CSF (Croatia); RPF (Cyprus); SENESCYT (Ecuador); MoER, ERC IUT, and ERDF (Estonia); Academy of Finland, MEC, and HIP (Finland); CEA and CNRS/IN2P3 (France); BMBF, DFG, and HGF (Germany); GSRT (Greece); OTKA and NIH (Hungary); DAE and DST (India); IPM (Iran); SFI (Ireland); INFN (Italy); MSIP and NRF (Republic of Korea); LAS (Lithuania); MOE and UM (Malaysia); BUAP, CINVESTAV, CONACYT, LNS, SEP, and UASLP-FAI (Mexico); MBIE (New Zealand); PAEC (Pakistan); MSHE and NSC (Poland); FCT (Portugal); JINR (Dubna); MON, RosAtom, RAS, and RFBR (Russia); MESTD (Serbia); SEIDI and CPAN (Spain); Swiss Funding Agencies (Switzerland); MST (Taipei); ThEPCenter, IPST, STAR, and NSTDA (Thailand); TUBITAK and TAEK (Turkey); NASU and SFFR (Ukraine); STFC (United Kingdom); DOE and NSF (USA).

\hyphenation{Rachada-pisek} Individuals have received support from the Marie-Curie programme and the European Research Council and EPLANET (European Union); the Leventis Foundation; the A. P. Sloan Foundation; the Alexander von Humboldt Foundation; the Belgian Federal Science Policy Office; the Fonds pour la Formation \`a la Recherche dans l'Industrie et dans l'Agriculture (FRIA-Belgium); the Agentschap voor Innovatie door Wetenschap en Technologie (IWT-Belgium); the Ministry of Education, Youth and Sports (MEYS) of the Czech Republic; the Council of Science and Industrial Research, India; the HOMING PLUS programme of the Foundation for Polish Science, cofinanced from European Union, Regional Development Fund, the Mobility Plus programme of the Ministry of Science and Higher Education, the National Science Center (Poland), contracts Harmonia 2014/14/M/ST2/00428, Opus 2013/11/B/ST2/04202, 2014/13/B/ST2/02543 and 2014/15/B/ST2/03998, Sonata-bis 2012/07/E/ST2/01406; the Thalis and Aristeia programmes cofinanced by EU-ESF and the Greek NSRF; the National Priorities Research Program by Qatar National Research Fund; the Programa Clar\'in-COFUND del Principado de Asturias; the Rachadapisek Sompot Fund for Postdoctoral Fellowship, Chulalongkorn University and the Chulalongkorn Academic into Its 2nd Century Project Advancement Project (Thailand); and the Welch Foundation, contract C-1845.

\end{acknowledgments}

\ifthenelse{\boolean{cms@external}}{}{\clearpage}

\bibliography{auto_generated}

\cleardoublepage \appendix\section{The CMS Collaboration \label{app:collab}}\begin{sloppypar}\hyphenpenalty=5000\widowpenalty=500\clubpenalty=5000\textbf{Yerevan Physics Institute,  Yerevan,  Armenia}\\*[0pt]
S.~Chatrchyan, V.~Khachatryan, A.M.~Sirunyan, A.~Tumasyan
\vskip\cmsinstskip
\textbf{Institut f\"{u}r Hochenergiephysik,  Wien,  Austria}\\*[0pt]
W.~Adam, T.~Bergauer, M.~Dragicevic, J.~Er\"{o}, C.~Fabjan\cmsAuthorMark{1}, M.~Friedl, R.~Fr\"{u}hwirth\cmsAuthorMark{1}, V.M.~Ghete, C.~Hartl, N.~H\"{o}rmann, J.~Hrubec, M.~Jeitler\cmsAuthorMark{1}, W.~Kiesenhofer, V.~Kn\"{u}nz, M.~Krammer\cmsAuthorMark{1}, I.~Kr\"{a}tschmer, D.~Liko, I.~Mikulec, D.~Rabady\cmsAuthorMark{2}, B.~Rahbaran, H.~Rohringer, R.~Sch\"{o}fbeck, J.~Strauss, A.~Taurok, W.~Treberer-Treberspurg, W.~Waltenberger, C.-E.~Wulz\cmsAuthorMark{1}
% ...
\vskip\cmsinstskip
% ...
1:~~Also at Vienna University of Technology, Vienna, Austria\\
2:~~Also at CERN, European Organization for Nuclear Research, Geneva, Switzerland\\
% ...

\end{sloppypar}
\end{document}